\documentclass[12pt]{article}
\usepackage{epsfig}
\usepackage{graphicx}
\usepackage{a4}
\usepackage{latexsym}
\usepackage{cite}

\usepackage{pslatex}
\usepackage[latin1]{inputenc}
\usepackage[T1]{fontenc}

\newcommand{\beq}{\begin{equation}}
\newcommand{\eeq}{\end{equation}}
\newcommand{\bea}{\begin{eqnarray}}
\newcommand{\eea}{\end{eqnarray}}
\newcommand{\nn}{\nonumber}

\newcommand{\MSb}{$\overline{\mbox{MS}}$}
\newcommand{\as}{\alpha_{\rm s}}
\newcommand{\ars}{a_{\rm s}}
\newcommand{\ep}{\epsilon}

\def\z#1{{\zeta_{#1}}}
\def\zs{{\zeta_{2}^{\,2}}}
\def\zt{{\zeta_{2}^{\,3}}}
\def\zf{{\zeta_{2}^{\,4}}}
\def\zzs{{\zeta_{3}^{\,2}}}
\def\ca{{C^{}_A}}
\def\cas{{C^{\,2}_A}}
\def\cat{{C^{\,3}_A}}
\def\cf{{C^{}_F}}
\def\cfs{{C^{\, 2}_F}}

\def\nf{{n^{}_{\! f}}}
\def\nsq{{n^{\,2}_{\! f}}}

\begin{document}
\setlength{\parskip}{0.2cm}
\setlength{\baselineskip}{0.56cm}

\begin{titlepage}
\noindent
DESY 05-138, SFB/CPP-05-34 \hfill {\tt hep-ph/0508055}\\
DCPT/05/88, $\,\,$IPPP/05/44\\
NIKHEF 05-013 \\[1mm]
August 2005 \\
\vspace{1.5cm}
\begin{center}
\Large
{\bf Three-Loop Results for Quark and Gluon Form Factors} \\
\vspace{2.0cm}
\large
S. Moch$^{\, a}$, J.A.M. Vermaseren$^{\, b}$ and A. Vogt$^{\, c}$\\
\vspace{1.2cm}
\normalsize
{\it $^a$Deutsches Elektronensynchrotron DESY \\
\vspace{0.1cm}
Platanenallee 6, D--15735 Zeuthen, Germany}\\
\vspace{0.5cm}
{\it $^b$NIKHEF Theory Group \\
\vspace{0.1cm}
Kruislaan 409, 1098 SJ Amsterdam, The Netherlands} \\
\vspace{0.5cm}
{\it $^c$IPPP, Department of Physics, University of Durham\\
\vspace{0.1cm}
South Road, Durham DH1 3LE, United Kingdom}\\
\vfill
\large
{\bf Abstract}
\vspace{-0.2cm}
\end{center}
We study the photon-quark-quark and Higgs-gluon-gluon form factors for on-shell
massless quarks and gluons in perturbative QCD. Previous third-order results
for the quark case are extended by calculating the fermion-loop contributions
up to the finite terms in dimensional regularization. For the gluon case the
complete set of infrared poles at three loops is derived. Using the 
exponentiation of the form factor, the latter results
are employed to extract a function entering the infrared factorization of
general third-order amplitudes. We evaluate the infrared finite absolute ratio 
of the time-like and space-like gluon form factors up to the fourth order in 
the strong coupling constant. The result supports previous indications that the
perturbative expansion of the Higgs boson production rate at the LHC is under 
control.  
\vspace{1.0cm}
\end{titlepage}

 
\noindent
The form factors of quarks and gluons, i.e., the QCD corrections to the $qqX$
and $ggX$ vertices with a colour-neutral particle $X$, are quantities of 
considerable phenomenological and theoretical interest.
These three-point amplitudes represent important gauge invariant (if infrared
divergent) parts of the perturbative corrections to processes of utmost 
relevance, like the production of lepton pairs and the Higgs boson $H$ at 
proton colliders. Due to the exponentiation of the form factors \cite
{Collins:1989bt,Magnea:1990zb,Magnea:2000ss}, already the dimensionally 
regulated pole terms at order $\as^{\,n}$, combined with appropriate 
lower-order information, suffice for deriving the finite absolute ratios of the
time-like and space-like form factors to order $\as^{\, n+1}$, thus providing 
valuable information about the size of the higher-order corrections.
The form factors are also the simplest amplitudes to which the infrared 
factorization formulae of Refs.~\cite{Catani:1998bh,Sterman:2002qn} can be 
applied. Consequently explicit calculations of these quantities lead, through 
the resulting determinations of otherwise unspecified functions in these
formulae, to unambiguous predictions for the pole structure of higher-order 
amplitudes involving more partons.

In a recent article~\cite{Moch:2005xx} we have extracted all three-loop pole 
terms of the electromagnetic form factor ${\cal F}^{\!q} (\as,Q^2)$ of on-shell 
massless quarks from the calculation of the third-order coefficient function in 
deep-inelastic scattering \cite{Vermaseren:2005qc} and extended the resummation
of this form factor to the next-to-next-to-leading contributions. In this 
letter we present the corresponding results for the $Hgg$ gluon form factor 
${\cal F}^{\!g} (\as,Q^2)$. 
As a first step towards the complete three-loop computations, we moreover 
extend the fermion-loop ($\nf$) parts of ${\cal F}^{\!q}$ to the order $\ep^0$ 
in dimensional regularization with $D = 4 - 2\,\ep$. 

Since results on deep-inelastic scattering are our starting point, we first
address the space-like form factors. Thus the relevant amplitude for the 
$\gamma^{\,\ast}\! qq$ quark case is
\beq
\label{eq:ffq}
\Gamma_\mu  \: = \: {\rm i} e_{\rm q}\,
\bigl({\bar u}\, \gamma_{\mu\,} u\bigr)\, {\cal F}^{\! q} (\as,Q^2)\: ,
\eeq
where $e_{\rm q}$ represents the quark charge and $Q^2$ the virtuality of the 
photon. The corresponding $Hgg$ vertex is an effective interaction in the 
limit of a heavy top quark, 
\beq
\label{eq:LggH} 
   {\cal L}_{\rm eff} \: = \: - \frac{1}{4} \: C_H \: H \, 
   G^{\,a}_{\mu\nu} G^{\,a,\mu\nu} \: .
\eeq
Here $G^{\,a}_{\mu\nu}$ denotes the gluon field strength tensor, and the 
prefactor $C_H$ is determined by the top-quark loop including all QCD 
corrections. Neither these corrections nor the renormalization of 
$ G^{\,a}_{\mu\nu} G^{\,a,\mu\nu}$ are relevant to the present study, hence we 
refer the reader to Refs.~\cite{Harlander:2000xx,Ravindran:2004xx} for details.

Besides the electromagnetic case utilized in Ref.~\cite{Moch:2005xx}, we have
also calculated the third-order corrections to deep-inelastic scattering by
exchange of a scalar interacting with gluons according to Eq.~(\ref{eq:LggH}).
The inclusion of this (experimentally entirely irrelevant) process was required
for obtaining the full set of splitting functions governing the evolution of
the parton distributions at the next-to-next-to-leading order~\cite
{Moch:2004pa,Vogt:2004mw}. All terms up to $\ep^0$ were consistently kept also 
in this part of the calculation. Hence we can repeat the procedure discussed
in Ref.~\cite{Moch:2005xx} for ${\cal F}^{\! g} (\as,Q^2)$, and derive the 
complete set of pole terms, $\ep^{-6} \ldots \ep^{-1}$, at three loops.
 
In order to access the finite ($\ep^0$) parts of the form factors in this
approach, the three-loop computations of deep-inelastic scattering need to be
extended to order $\ep$. This is not an option for the complete calculation 
with its more than 100$\,$000 tabulated integrals. However, only very few 
genuine three-loop diagrams remain if we confine ourselves to the $\nf$ 
contributions to the quark form factor (these diagrams can be found in 
Ref.~\cite {Moch:2002sn}). As neither of those diagrams is of a particularly 
difficult type, we were able to carry out the required extensions for this 
subset.

We present our results in terms of the expansion coefficients ${\cal F}_{l}^
{\! p}$ of the bare (unrenormalized) form factors ($\,p$ = $q,\: g\,$)
\beq
\label{eq:ffexp}
  {\cal F}_{\rm{b}}^p(\as^{\,\rm{b}},Q^2) \: = \:
  1 + \sum\limits_{l=1}^\infty\, \: \Big(\ars^{\rm{b}} \Big)^l \,
  \Bigg({Q^2 \over \mu^2}\Bigg)^{\! -l\epsilon} {\cal F}_l^{\! p} 
  \:\: , \quad \ars \: \equiv \: \frac{\as}{4\pi} \:\: .
\eeq
The fermionic contributions to the quark form factor are given by
\bea
  {\cal F}_{1,\nf}^q & = & 
       0
\label{eq:ff1q-nf}
\: ,  \\[2mm]
  {\cal F}_{2,\nf}^q & = & 
    \cf \* \nf \* \Bigg\{
         {1 \over 3 \* \epsilon^3}
       + {14 \over 9 \* \epsilon^2}
       + {1 \over \epsilon} \* \Bigg(
            {353 \over 54} 
          + {1 \over 3} \* \z2
          \Bigg)
          + {7541 \over 324}
          + {14 \over 9} \* \z2
          - {26 \over 9} \* \z3
\nonumber\\
& &\mbox{}
       + \epsilon \* \left(
            {150125 \over 1944}
          + {353 \over 54} \* \z2
          - {364 \over 27} \* \z3
          - {41 \over 30} \* \zs
          \right) 
\nonumber\\
& &\mbox{}
       + \epsilon^2 \* \Bigg(
            {2877653 \over 11664}
          - {26 \over 9} \* \z2 \* \z3
          + {7541 \over 324} \* \z2
          - {4589 \over 81} \* \z3
          - {287 \over 45} \* \zs
          - {242 \over 15} \* \z5
          \Bigg)
          \Bigg\}
\label{eq:ff2q-nf}
\: ,  \\[3mm]
  {\cal F}_{3,\nf}^q & = & 
    \cfs \* \nf \* \Bigg\{
       - {2 \over 3 \* \epsilon^5}
       - {37 \over 9 \* \epsilon^4}
       + {1 \over \epsilon^3}  \*  \Bigg(
          - {545 \over 27}
          - {1 \over 3} \* \z2
          \Bigg)
       + {1 \over \epsilon^2}  \*  \Bigg(
          - {6499 \over 81}
          - {133 \over 18} \* \z2
\nonumber\\
& &\mbox{}
          + {146 \over 9} \* \z3
          \Bigg)
       + {1 \over \epsilon}  \*  \Bigg(
          - {138865 \over 486}
          - {2849 \over 54} \* \z2
          + {2557 \over 27} \* \z3
          + {337 \over 36} \* \zs
          \Bigg)
\nonumber\\
& &\mbox{}
       - {2732173 \over 2916}
          - {45235 \over 162} \* \z2
          + {51005 \over 81} \* \z3  
          + {8149 \over 216} \* \zs
          - {343 \over 9} \* \z2 \* \z3
          + {278  \over 45} \* \z5 
          \Bigg\}
\nonumber\\
& &\mbox{}
   + \cf \* \ca \* \nf \* \Bigg\{
         {88 \over 81 \* \epsilon^4}
       + {1 \over \epsilon^3}  \*  \left(
            {2254 \over 243}
          - {16 \over 27} \* \z2
          \right)
       + {1 \over \epsilon^2}  \*  \Bigg(
            {13679 \over 243}
          + {316 \over 81} \* \z2
\nonumber\\
& &\mbox{}
          - {256 \over 27} \* \z3
          \Bigg)
       + {1 \over \epsilon}  \*  \left(
            {623987 \over 2187}
          + {11027 \over 243} \* \z2
          - {6436 \over 81} \* \z3
          - {44 \over 5} \* \zs
          \right)
       + {8560052 \over 6561}
\nonumber\\
& &\mbox{}
          + {442961 \over 1458} \* \z2
          - {45074 \over 81} \* \z3  
          - {1093 \over 27} \* \zs
          + {368 \over 9} \* \z2 \* \z3
          - {208  \over 3} \* \z5 
          \Bigg\}
\nonumber\\
& &\mbox{}
  + \cf \* \nsq \* \Bigg\{
       - {8 \over 81 \* \epsilon^4}
       - {188 \over 243 \* \epsilon^3}
       + {1 \over \epsilon^2}  \*  \left(
          - {124 \over 27}
          - {4 \over 9} \* \z2
          \right)
       + {1 \over \epsilon}  \*  \Bigg(
          - {49900 \over 2187}
          - {94 \over 27} \* \z2
\nonumber\\
& &\mbox{}
          + {136 \over 81} \* \z3
          \Bigg)
       - { 677716 \over 6561}
          - {62 \over 3} \* \z2
          + {3196 \over 243} \* \z3  
          - {83 \over 135} \* \zs
          \Bigg\}
\label{eq:ff3q-nf}
\: .
\eea
Here $\nf$ stands for the number of effectively massless quark flavours, $C_F$ 
and $C_A$ are the usual QCD colour factors, $C_F= 4/3$ and $C_A= 3$, and the 
values of Riemann's zeta function are denoted by $\zeta_n$.
The $\ep^1$ and $\ep^2$ terms in Eq.~(\ref{eq:ff2q-nf}) and the corresponding
non-fermionic contributions~\cite{Moch:2005xx} have recently been confirmed in
Ref.~\cite{Gehrmann:2005pd}, where exact expression for both two-loop form
factors were derived. 
The finite ($\ep^0$) contributions in Eq.~(\ref{eq:ff3q-nf}) represent a new 
result of the present letter.

The first three expansion coefficients (\ref{eq:ffexp}) of the unrenormalized
gluon form factor read
\bea
  {\cal F}_1^g &\! = & 
  \ca \* \Bigg\{
       - {2 \over \epsilon^2}
       + \z2
       + \epsilon  \*  \left(
          - 2
          + {14 \over 3} \* \z3
          \right)
       + \epsilon^2  \*  \left(
          - 6
          + {47 \over 20} \* \zs
          \right)
       + \epsilon^3  \*  \Bigg(
          - 14
          + \z2
\nonumber\\
& &\mbox{}
          - {7 \over 3} \* \z2 \* \z3
          + {62 \over 5} \* \z5
          \Bigg)
       + \epsilon^4 \*  \Bigg(
          - 30
          + 3 \* \z2
          + {14 \over 3} \* \z3
          + {949 \over 280} \* \zt
          - {49 \over 9} \* \zzs
          \Bigg)
          \Bigg\}
\label{eq:ff1ggH}
\: ,  \\[2mm]
\label{eq:ff2ggH}
  {\cal F}_2^g &\! = & 
  \cas \* \Bigg\{
         {2 \over \epsilon^4}
       - {11 \over 6 \* \epsilon^3}
       + {1 \over \epsilon^2} \* \left(
          - {67 \over 18}
          - \z2
          \right)
       + {1 \over \epsilon} \*  \left(
            {68 \over 27}
          + {11 \over 2} \* \z2
          - {25 \over 3} \* \z3
          \right)
       + {5861 \over 162} 
       + {67 \over 6} \* \z2 
\nonumber\\
& &\mbox{}
       + {11 \over 9} \* \z3 
       - {21 \over 5} \* \zs 
       + \epsilon \*  \left(
            {158201 \over 972}
          + {106 \over 9} \* \z2
          - {1139 \over 27} \* \z3
          - {77 \over 60} \* \zs
          + {23 \over 3} \* \z2 \* \z3
          + {71 \over 5} \* \z5
          \right)
\nonumber\\
& &\mbox{}
       + \epsilon^2 \* \Bigg(
            {3484193 \over 5832}
          + {481 \over 54} \* \z2
          - {26218 \over 81} \* \z3
          - {1943 \over 60} \* \zs
          - {55 \over 3} \* \z2 \* \z3
          + {341 \over 15} \* \z5
          + {2313 \over 70} \* \zt
\nonumber\\
& &\mbox{}
          + {901 \over 9} \* \zzs
          \Bigg)
          \Bigg\}
 \:\:  + \:\: \ca \* \nf \* \Bigg\{
         {1 \over 3 \* \epsilon^3}
       + {5 \over 9 \* \epsilon^2}
       + {1 \over \epsilon} \* \left(
          - {26 \over 27}
          - \z2
          \right)
          - {808 \over 81}
          - {5 \over 3} \* \z2
          - {74 \over 9} \* \z3
\nonumber\\
& &\mbox{}
       + \epsilon \* \Bigg(
          - {23131 \over 486}
          - {16 \over 9} \* \z2
          - {604 \over 27} \* \z3
          - {51 \over 10} \* \zs
          \Bigg)
       + \epsilon^2 \* \Bigg(
          - {540805 \over 2916}
          + {28 \over 27} \* \z2
          - {3962 \over 81} \* \z3
\nonumber\\
& &\mbox{}
          - {257 \over 18} \* \zs
          + {50 \over 3} \* \z2 \* \z3
          - {542 \over 15} \* \z5
          \Bigg)
          \Bigg\}
 \:\: + \:\: \cf \* \nf \* \Bigg\{
       - {1 \over \epsilon}
          - {67 \over 6}
          + 8 \* \z3
       + \epsilon \* \Bigg(
          - {2027 \over 36}
\nonumber\\
& &\mbox{}
          + {7 \over 3} \* \z2
          + {92 \over 3} \* \z3
          + {16 \over 3} \* \zs
          \Bigg) 
       + \epsilon^2 \* \Bigg(
          - {47491 \over 216}
          + {209 \over 18} \* \z2
          + {1124 \over 9} \* \z3
          + {184 \over 9} \* \zs
\nonumber\\
& &\mbox{}
          - {40 \over 3} \* \z2 \* \z3
          + 32 \* \z5
          \Bigg)
          \Bigg\}
\: ,\quad \nonumber\\[2mm]
  {\cal F}_3^g &\! = & 
  \cat \* \Bigg\{
       - {4 \over 3 \* \epsilon^6}
       + {11 \over 3 \* \epsilon^5}
       + {361 \over 81 \* \epsilon^4}
       + {1 \over \epsilon^3}  \*  \left(
          - {3506 \over 243}
          - {517 \over 54} \* \z2
          + {22 \over 3} \* \z3
          \right)
       + {1 \over \epsilon^2}  \*  \Bigg(
          - {17741 \over 243}
\nonumber\\
& &\mbox{}
          + {481 \over 162} \* \z2
          - {209 \over 27}  \* \z3
          + {247 \over 90} \* \zs
          \Bigg)
       + {1 \over \epsilon}  \*  \Bigg(
          - {145219 \over 2187}
          + {20329 \over 243} \* \z2
          + {241 \over 9} \* \z3
          - {3751 \over 360} \* \zs
\nonumber\\
& &\mbox{}
          - {85 \over 9} \* \z2 \* \z3
          - {878 \over 15} \* \z5
          \Bigg)
          \Bigg\}
 \:\: + \:\: \cas \* \nf \* \Bigg\{
       - {2 \over 3 \* \epsilon^5}
       - {2 \over 81 \* \epsilon^4}
       + {1 \over \epsilon^3}  \*  \left(
            {1534 \over 243}
          + {47 \over 27} \* \z2
          \right)
\nonumber\\
& &\mbox{}
       + {1 \over \epsilon^2}  \*  \Bigg(
            {4280 \over 243}
          - {425 \over 81} \* \z2
          + {518 \over 27} \* \z3
          \Bigg)
       + {1 \over \epsilon}  \*  \Bigg(
          - {92449 \over 2187}
          - {7561 \over 243} \* \z2
          + {1022 \over 81} \* \z3
          + {2453 \over 180} \* \zs
          \Bigg)
          \Bigg\}
\nonumber\\
& &\mbox{}
+ \ca \* \cf \* \nf \* \Bigg\{
         {20 \over 9 \* \epsilon^3}
       + {1 \over \epsilon^2}  \*  \left(
            {526 \over 27}
          - {160 \over 9} \* \z3
          \right)
       + {1 \over \epsilon}  \*  \Bigg(
            {2783 \over 81}
          - {22 \over 3} \* \z2
          - {224 \over 27} \* \z3
          - {176 \over 15} \* \zs
          \Bigg)
          \Bigg\}
\nonumber\\
& &\mbox{}
+ \cfs \* \nf \* \Bigg\{ 
            {2 \over 3 \* \epsilon} 
          \Bigg\}
 \:\: + \:\: \ca \* \nsq \* \Bigg\{ 
       - {8 \over 81 \* \epsilon^4}
       - {80 \over 243 \* \epsilon^3}
       + {1 \over \epsilon^2}  \*  \left(
            {8 \over 9}
          + {20 \over 27} \* \z2
          \right)
       + {1 \over \epsilon}  \*  \Bigg(
            {34097 \over 2187}
          + {200 \over 81} \* \z2
\nonumber\\
& &\mbox{}
          + {664 \over 81} \* \z3
          \Bigg)
          \Bigg\}
 \:\: + \:\: \cf \* \nsq \* \Bigg\{
         {8 \over 9 \* \epsilon^2}
       + {1 \over \epsilon}  \*  \Bigg(
            {424 \over 27}
          - {32 \over 3} \* \z3
          \Bigg)
          \Bigg\}
\label{eq:ff3ggH}
\: .
\eea
As in Eq.~(\ref{eq:ff2q-nf}), the one- and two-loop quantities 
(\ref{eq:ff1ggH}) and (\ref{eq:ff2ggH}) are written down to the accuracy
in $\ep$ required for the extraction of the three-loop form factors to order
$\ep^0$. The terms up this order in Eq.~(\ref{eq:ff2ggH}) have been 
obtained before in Refs.~\cite{Harlander:2000xx,Ravindran:2004xx}. Our 
corresponding coefficients of $\ep^1$ and $\ep^2$ agree with the recent 
all-order expression of Ref.~\cite{Gehrmann:2005pd}. 
Eq.~(\ref{eq:ff3ggH}), of which we will present an independent check below,
is the main new result of this article. 
 
In the context of Refs.~\cite{Kotikov:2004er,Bern:2005iz} it is interesting to
note that the terms of highest transcendentality, i.e, the coefficients of
$\zeta_n$ and $\zeta_i\,\zeta_j$ with $\,i+j=n$, in the $\ep^{-2l+n}$ 
contributions to ${\cal F}_l$ agree between the quark and gluon form factors 
for the Super-Yang-Mills case $C_A = C_F = n_c$. In fact, up to two loops this
can be readily shown to all orders in $\ep$ by expanding the prefactors of the
master integrals in Eqs.~(8) -- (11) of Ref.~\cite{Gehrmann:2005pd}.

The exponentiation of the form factors is performed for the (coupling-constant)
renormalized quantities. These are obtained from Eqs.~(\ref{eq:ffexp}) --
(\ref{eq:ff3ggH}) by replacing the bare coupling $\ars^{\,b}$ by its 
renormalized counterpart $\ars$ according to
\beq
\label{eq:asren}
\ars^{\,\rm b} \: = \: \ars  \left\{ 1 - {\beta_0 \over \epsilon} \ars\,
+ \left({\beta_0^2 \over \epsilon^2}
  - {1 \over 2} {\beta_1 \over \epsilon}\right) \ars^{\,2}\,
- \left({\beta_0^3 \over \epsilon^3}
   - {7 \over 6} {\beta_1 \beta_0 \over \epsilon^2}
   + {1 \over 3} {\beta_2 \over \epsilon}\right) \ars^{\,3} \right\} \: ,
\eeq
where $\beta_{\,\rm i}$ are the usual expansion coefficients of the beta 
function of QCD, $\beta_0 = \frac{11}{3}\,\ca - \frac{2}{3}\,\nf$ etc.
Note that, unlike Refs.~\cite{Ravindran:2004xx,Gehrmann:2005pd}, we do not
include the multiplicative renormalization of $G^{\,a}_{\mu\nu} G^{\,a,\mu\nu}$
in Eq.~(\ref{eq:LggH}) into the definition of the (renormalized) gluon form 
factor.

The exponentiation of the (coupling-constant) renormalized form factors can be
written as~\cite{Magnea:1990zb,Magnea:2000ss}
\bea
\label{eq:ffres}
\ln {\cal F}\! \left(\as, {Q^2 \over \mu^2}, \epsilon\right) &\! = &
{1 \over 2} \int\limits_0^{Q^2/\mu^2}\, {d \xi \over \xi} \Bigg(
  K(\as,\epsilon) + G(1,{\bar a}(\xi,a_s,\epsilon),\epsilon)
  + \int\limits_\xi^1\, {d \lambda \over \lambda}
  A({\bar a}(\lambda,\ars,\epsilon)) \Bigg) \: . \quad 
\eea
Here $K^p(\as,\ep)$ are scale-independent counter-term functions consisting of 
a series of poles in $\ep$ in the \MSb\ scheme. These functions can be 
determined recursively from a renormalization group equation (see, e.g., 
Ref.~\cite{Magnea:2000ss}) in terms of the cusp anomalous dimensions $A^p$
\cite{Korchemsky:1989si}. 
The functions $G^p$, on the other hand, can be expanded in non-negative powers 
of $\ep$ at all orders of $\as$. Finally $\bar{a}$ is the running coupling in 
$D$ dimensions, see Refs.~\cite{Contopanagos:1997nh,Moch:2005xx}.
After expansion in $\as$, the integrals in Eq.~(\ref{eq:ffres}) can be solved
using algorithms for the evaluation of nested sums \cite{Vermaseren:1998uu,%
Moch:2001zr}, see Ref.~\cite{Moch:2005xx}. 
Transforming back to the unrenormalized expansion coefficients
${\cal F}_i^{\! p}$ in Eq.~(\ref{eq:ffexp}) the results read
\bea
  {\cal F}_1^{} & = & 
          - {1 \over 2} \* {1 \over \epsilon^2} \* A_1
          - {1 \over 2} \* {1 \over \epsilon} \* G_1
\label{eq:ff1loop}
\, ,  \\[1mm]
  {\cal F}_2^{} & = & 
            {1 \over 8} \* {1 \over \epsilon^4} \* A_1^2
          + {1 \over 8} \* {1 \over \epsilon^3} \* A_1 \* ( 
            2 \* G_1
          - \beta_0
          )
          + {1 \over 8} \* {1 \over \epsilon^2} \* (
            G_1^2 
          - A_2 
          - 2 \* \beta_0 \* G_1
          )
          - {1 \over 4} \* {1 \over \epsilon} \* G_2
\label{eq:ff2loop}
\, ,  \\[1mm]
  {\cal F}_3^{} & = & 
          - {1 \over 48} \* {1 \over \epsilon^6} \* A_1^3
          - {1 \over 16} \* {1 \over \epsilon^5} \* A_1^2 \* ( 
            G_1 
          - \beta_0
          )
          - {1 \over 144} \* {1 \over \epsilon^4} \* A_1 \* (
            9 \* G_1^2 
          - 9 \* A_2 
          - 27 \* \beta_0 \* G_1 
          + 8 \* \beta_0^2
          )
\nonumber\\
& &\mbox{}
          - {1 \over 144} \* {1 \over \epsilon^3} \* ( 
            3 \* G_1^3 
          - 9 \* A_2 \* G_1 
          - 18 \* A_1 \* G_2 
          + 4 \* \beta_1 \* A_1 
          - 18 \* \beta_0 \* G_1^2 
          + 16 \* \beta_0 \* A_2 
          + 24 \* \beta_0^2 \* G_1
          )
\nonumber\\
& &\mbox{}
          + {1 \over 72} \* {1 \over \epsilon^2} \* (
            9 \* G_1 \* G_2 
          - 4 \* A_3 
          - 6 \* \beta_1 \* G_1 
          - 24 \* \beta_0 \* G_2
          )
          - {1 \over 6} \* {1 \over \epsilon} \* G_3
\label{eq:ff3loop}
\, .
\eea
The corresponding expression for the four-loop form factors can be found in 
Ref.~\cite{Moch:2005xx}.

From now on we confine ourselves to the gluon form factor ${\cal F}^{\! g}$.
The corresponding anomalous dimensions $A^g$ are related by a factor $\ca/\cf$
to the quark quantities $A^q$~\cite{Korchemsky:1989si} which are known to order
$\as^{\,3}$ \cite{Moch:2004pa,Kodaira:1982nh}. This relation has been verified 
by the explicit calculation of Ref.~\cite{Vogt:2004mw}. As always using the
expansion parameter $\ars = \as/(4\pi)$, the available coefficients read    
\bea
  A_1^g  & = &  4 \,\ca \: , 
\label{eq:a1}
\\[1mm]  
  A_2^g  & = &  8 \,\cas \left( \frac{67}{18} - \z2 \right)
            + 8 \,\ca \nf \left( - \frac{5}{9} \right) \: ,
\label{eq:a2}
\\[1mm]
  A_3^g & = &
     16\, \cat \, \left( \frac{245}{24} - \frac{67}{9}\: \z2
         + \frac{11}{6}\:\z3 + \frac{11}{5}\:\zs \right)
   +  16\, \ca \cf \nf\, \left( -  \frac{55}{24}  + 2\:\z3 \right)
   \quad \nn \\ & & \mbox{}
   +  16\, \cas \nf\, \left( - \frac{209}{108}
         + \frac{10}{9}\:\z2 - \frac{7}{3}\:\z3 \right)
   \, +\,  16\, \ca \nsq \left( - \frac{1}{27}\,\right) \: .
\label{eq:a3}
\eea
Beyond the third order only the (small) leading-$\nf$ contributions are known
~\cite{Gracey:1994nn}.
 
Using these coefficients, the $l$-loop pole terms $\ep^{-2l} \ldots\,\ep^{-2}$ 
in Eqs.~(\ref{eq:ff1loop}) -- (\ref{eq:ff3loop}) can be predicted from 
lower-order information, thus providing a strong check of either the 
exponentiation formula or the explicit $l$-th order calculation. Our results 
pass this check, thus Eqs.~(\ref{eq:ff1ggH}) -- (\ref{eq:ff3ggH}) can be 
employed to recursively derive the resummation functions $G_i^{\, g}$ for 
$i= 1,\,2,$ and 3 to the respective fifth, third and zeroth order in $\ep$,
yielding
\bea
\label{eq:g1g}
  G_1^{\,g} & = &
          \epsilon \* \ca \* (- 2 \* \z2)
        + \epsilon^2 \* \ca \* \Bigg( 4 - {28 \over 3} \* \z3 \Bigg)
        + \epsilon^3 \* \ca \* \Bigg( 12 - {47 \over 10} \* \zs \Bigg)
        + \epsilon^4 \* \ca \* \Bigg( 28  - 2 \* \z2
\nonumber \\
& & \mbox{}
        + {14 \over 3} \* \z2 \* \z3 - {124 \over 5} \* \z5 \Bigg)
        + \epsilon^5 \* \ca \* \Bigg(
          60 - 6 \* \z2 - {28 \over 3} \* \z3 
        - {949 \over 140} \* \zt + {98 \over 9} \* \zzs 
        \Bigg)
\: ,
\\[3mm]
\label{eq:g2g}
  G_2^{\,g} & = & 
 \cas \* \Bigg(
         {160 \over 27}
       - {44 \over 3} \* \z2
       - 4 \* \z3
       \Bigg)
 + \ca \* \nf \* \Bigg(
         {104 \over 27} 
       + {8 \over 3} \* \z2 
       \Bigg)
 + 4 \* \cf \* \nf 
\nonumber \\
& & \mbox{}
 + \ep \* \cas \* \Bigg(
        - {9022 \over 81} 
        - {134 \over 3} \* \z2 
        + {88 \over 3} \* \z3
        \Bigg) 
 + \ep \* \ca \* \nf \* \Bigg(
          {3448 \over 81} 
        + {20 \over 3} \* \z2 
        + {80 \over 3} \* \z3
        \Bigg) 
\nonumber \\
& & \mbox{}
 + \ep \* \cf \* \nf \* \Bigg(
          {134 \over 3} 
        - 32 \* \z3
        \Bigg) 
 + \ep^2 \* \cas \* \Bigg( 
        - {141677 \over 243} 
        - {568 \over 9} \* \z2 
        + {4556 \over 27} \* \z3
        + {671 \over 30} \* \zs
\nonumber \\
& & \mbox{}
        + {20 \over 3} \* \z2 \* \z3
        -156 \* \z5
        \Bigg) 
 + \ep^2 \* \ca \* \nf \* \Bigg( 
          {48206 \over 243} 
        + {64 \over 9} \* \z2 
        + {2416 \over 27} \* \z3
        + {259 \over 15} \* \zs
        \Bigg) 
\nonumber \\
& & \mbox{}
 + \ep^2 \* \cf \* \nf \* \Bigg( 
          {2027 \over 9} 
        - {28 \over 3} \* \z2 
        - {368 \over 3} \* \z3
        - {64 \over 3} \* \zs
        \Bigg) 
 + \ep^3 \* \cas  \*  \Bigg(  
        - {3272297 \over 1458} 
\nonumber \\
& & \mbox{} 
        - {2060 \over 27} \* \z2 
        + {98824 \over 81} \* \z3
        + {1943 \over 15} \* \zs
        + {506 \over 9} \* \z2 \* \z3
        - {5246 \over 35} \* \zt
        - {940 \over 3} \* \zzs
        \Bigg) 
\nonumber \\
& & \mbox{} 
 + \ep^3 \* \ca \* \nf  \*  \Bigg( 
          {554413 \over 729} 
        - {148 \over 27} \* \z2 
        + {15848 \over 81} \* \z3 
        + {514 \over 9} \* \zs 
	   - {572 \over 9} \* \z2 \* \z3
	   + 128 \* \z5
        \Bigg) 
\nonumber \\
& & \mbox{} 
 + \ep^3 \* \cf \* \nf  \*  \Bigg(   
          {47491 \over 54} 
        - {418 \over 9} \* \z2 
        - {4496 \over 9} \* \z3 
        - {736 \over 9} \* \zs 
	   + {160 \over 3} \* \z2 \* \z3
	   - 128 \* \z5
        \Bigg) 
\: ,
\\[3mm]
\label{eq:g3g}
  G_3^{\,g} & = & 
   \cat \*  \Bigg(
        - {373975 \over 729}
        - {27320 \over 81} \* \z2
        + {4096 \over 27} \* \z3
        + {1276 \over 15} \* \zs
        + {80 \over 3} \* \z2 \* \z3
        + 32 \* \z5
        \Bigg)
\nonumber \\
& & \mbox{}
 + \cas \* \nf \* \Bigg( 
          {266072 \over 729}
        + {7328 \over 81} \* \z2
        + {56 \over 9} \* \z3
        - {328 \over 15} \* \zs
        \Bigg)
 + \ca \* \cf \* \nf \* \Bigg( 
          {3833 \over 27}
        + 8 \* \z2
\nonumber \\
& & \mbox{}
        - {752 \over 9} \* \z3
        + {32 \over 5} \* \zs
        \Bigg)
 - 4 \* \cfs \* \nf
 + \ca \* \nsq \* \Bigg( 
        - {28114 \over 729}
        - {160 \over 27} \* \z2
        - {256 \over 27} \* \z3
        \Bigg)
\nonumber \\
& & \mbox{}
 + \cf \* \nsq \* \Bigg( 
        - {104 \over 3}
        + {64 \over 3} \* \z3
        \Bigg)
\: .
\eea
Results analogous to the terms up to order $\ep$ in Eqs.~(\ref{eq:g1g}) and
(\ref{eq:g2g}) have been derived before, in a different notation, in Ref.\
\cite{Ravindran:2004xx}. The third-order coefficient (\ref{eq:g3g}) is a new 
result based on Eq.~(\ref{eq:ff3ggH}). Note that $G_3^{\,g}$ is one of the 
quantities which determine the infrared structure of QCD amplitudes in the 
framework of Refs.~\cite{Catani:1998bh,Sterman:2002qn}.

We now turn to the check of our main results (\ref{eq:ff3ggH}) and 
(\ref{eq:g3g}) announced above. Inspired by a key observation of Ref.~\cite
{Ravindran:2004xx}, we decompose the resummation functions $G_i^{\,p}$ ($p =
q,\:g$) according to 
\bea
\label{eq:gf}
  G_1^{\,p} & = & 2 \left( B_1^{\,p} - \:\delta_{pg} \beta_0 \right) \:\:
     + f_1^{\,p} + \ep \widetilde{G_1^{\,p}} 
\:\: , \nn \\[1mm]
  G_2^{\,p} & = & 2 \left( B_2^{\,p} - 2 \delta_{pg} \beta_1 \right)
     + f_2^{\,p} + \beta_0 \widetilde{G_1^{\,p}}(\ep\!=\!0) 
     + \ep \widetilde{G_2^{\,p}} 
\:\: , \\
  G_3^{\,p} & = & 2 \left( B_3^{\,p} - 3 \delta_{pg} \beta_2 \right)
     + f_3^{\,p} + \beta_1 \widetilde{G_1^{\,p}}(\ep\!=\!0) 
     + \beta_0 \Big[ \widetilde{G_2^{\,p}}(\ep\!=\!0) 
     - \beta_0\widetilde{\widetilde{G_1^{\,p}}}(\ep\!=\!0) \Big] 
     + \ep \widetilde{G}_3^{\,p} 
\nn
\eea
with
\beq
  \widetilde{F} \:\: = \:\: \ep^{-1} 
  \left[ \, F - F (\ep\!=\!0) \, \right]
\eeq
and $B_n^{\,p}$ denoting the coefficients of $\delta(1-x)$ in the $n$-loop 
diagonal \MSb\ splitting functions $P_{pp}^{(n-1)}(x)$ \cite{Moch:2004pa,%
Vogt:2004mw}. The crucial point of this decomposition is that the functions
$f_i^{\,p}$ are, like the cusp anomalous dimensions $A_n^{\,p}$, universal 
up to the factor $\ca/\cf$, i.e., $f_i^{\,g} = \ca/\cf \: f_i^{\,q}$.
The functions $f_i^{\,p}$ also exhibit the same maximally non-Abelian colour 
structure as the $A_n^{\,p}$~\cite{Korchemsky:1989si}, with
\bea
  f_1^{\,q} & = & 0 
\:\: , \\[2mm]
  f_2^{\,q} & = & 2 \cf \left\{ 
    - \beta_0 \z2 - \frac{56}{27}\, \nf + \ca \left( 
    \frac{404}{27} - 14 \z3 \right) \right\}
\:\: , \\[3mm]
  f_3^{\,q} & = & 
  \cf \cas \Bigg(
     {136781 \over 729} - {12650 \over 81}\,\z2 - {1316 \over 3}\, \z3
   + {352 \over 5}\,\zs + {176 \over 3}\,\z2\z3 + 192\,\z5
  \Bigg)
\nn \\[1mm] & & \mbox{}
+ \ca\cf\nf \Bigg(
   - {11842 \over 729} 
   + {2828 \over 81}\,\z2 + {728 \over 27}\, \z3 - {96 \over 5}\,\zs 
  \Bigg)
+ \cfs\nf \Bigg(
   - {1711 \over 27} 
\nn \\[1mm] & & \mbox{}
   + 4\,\z2 + {304 \over 9}\, \z3 + {32 \over 5}\,\zs 
  \Bigg)
+ \cf\nsq \Bigg(
   - {2080 \over 729} - {40 \over 27} \,\z2 + {112 \over 27}\, \z3
  \Bigg)
\:\: .
\eea
Consequently, once $f_3^{\,q}$ is known from the explicit calculation of the
three-loop quark form factor~\cite{Moch:2005xx}, the pole terms of the gluon
form factor can be derived from the results of Refs.~\cite{Moch:2004pa,%
Vogt:2004mw} and lower-order quantities. This procedure confirms 
Eq.~(\ref{eq:g3g}) and thus Eq.~(\ref{eq:ff3ggH}).
For applications to other cases, like the coupling of a preudoscalar Higgs
boson, it should be noted that the terms subtracted from $B_i^{\,g}$ in Eqs.\
(\ref{eq:gf}) are given by the renormalization of the operator in 
Eq.~(\ref{eq:LggH}).

As a first concrete application of our results we consider the absolute ratio
$|{\cal F}^{\! g}(q^2)/{\cal F}^{\! g}(-q^2)|$ of the renormalized time-like 
and space-like form factors. This quantity is infrared finite and directly 
enters the cross section for Higgs boson production in hadronic collisions.
In terms of the coefficients $A_i$ and $G_i(\ep\!=\!0)$ this ratio is given 
by \cite{Moch:2005xx} 
\bea
\label{eq:FFrat}
\left| {{\cal F}(q^2) \over {\cal F}(-q^2)} \right|^2  & = & 
          1 
        + \ars \* \{
            3 \* \z2 \* A_1
          \}
        + \as^{\,2} \* \biggl\{
            {9 \over 2} \* \zs \* A_1^2 
          + 3 \* \z2 \* (\beta_0 \* G_1 + A_2)
          \biggr\}
\nonumber \\[-1mm]
& & \mbox{}
        + \ars^{\,3} \* \biggl\{
            {9 \over 2} \* \zt \* A_1^3
          + 3 \* \zs \* A_1 \* (3 \* \beta_0 \* G_1 - \beta_0^2 + 3 \* A_2)
          + 3 \* \z2 \* (A_3 + \beta_1 \* G_1 + 2 \* \beta_0 \* G_2)
          \biggr\}
\nonumber \\[1mm]
& & \mbox{}
        + \ars^{\,4} \* \biggl\{
            {27 \over 8} \* \zf \* A_1^4
          + {9 \over 2} \* \zt  \* A_1^2 \* (3 \* \beta_0 \* G_1 
          - 2 \* \beta_0^2 + 3 \* A_2)
          + {3 \over 2} \* \zs \* (
            - 6 \* \beta_0^2 \* A_2 
            + 3 \* \beta_0^2 \* G_1^2 
\nonumber \\
& & \mbox{}\quad\quad
            + 3 \* A_2^2 
            + 12 \* \beta_0 \* A_1 \* G_2 
            + 6 \* A_1 \* A_3
            + 6 \* \beta_1 \* A_1 \* G_1 
            - 5 \* \beta_0 \* \beta_1 \* A_1 
            + 6 \* \beta_0 \* A_2 \* G_1 
\nonumber \\
& & \mbox{}\quad\quad
            - 6 \* \beta_0^3 \* G_1)
          + 3 \* \z2 \* (A_4 + \beta_2 \* G_1 + 3 \* \beta_0 \* G_3 
          + 2 \* \beta_1 \* G_2)
          \biggr\} \:\: + \:\: {\cal O} (\ars^{\,5})
\eea
for the couplings $\ars(q^2) = \ars(-q^2) = \ars$.
All terms contributing at the fourth order are now known, with the exception of
the small four-loop cusp anomalous dimension $A_4^g$. For the latter we can 
employ the [1/1] Pad\'e estimate of the quark case in Ref.~\cite{Moch:2005ba}, 
accordingly multiplied by $\ca/\cf$,
\beq
\label{eq:A4pade}
  A_4^g \; \approx \; 17660\: ,\:\: 9704\: ,\:\:  3949 \quad
  \mbox{for} \quad \nf \; = \; 3\: ,\:\: 4\: ,\:\: 5 \:\: .
\eeq
Below a conservative (but numerically irrelevant) uncertainty of 50\% is 
assigned to this estimate. 

Inserting the numerical values of Eqs.~(\ref{eq:a1}) -- (\ref{eq:g3g}) for 
$\nf=5$ quark flavours, we obtain
\beq
\label{eq:ffr-num}
\left| {{\cal F}^{\! g}(q^2) \over {\cal F}^{\! g}(-q^2)} \right|^2 \: = \:
           1 + 4.712\: \as + 13.69\: \as^{\,2}
             + 25.94\: \as^{\,3} + (36.65 \pm 0.35)\: \as^{\,4} \: .
\eeq
The contributions up to order $\as^2$ have been derived before in Ref.~\cite
{Harlander:2000xx}, where the factor 153.2 in the penultimate line of Eq.~(19)
should read 135.2. The low-order contributions to Eq.~(\ref{eq:ffr-num}) are
larger than for the corresponding quark quantity in Ref.~\cite{Moch:2005xx}.
On the other hand the growth of the coefficients is slower in the present 
case, surprisingly suggesting a more benign higher-order behaviour. 
As for Higgs boson production couplings of the order $\as \approx 0.1$ are 
relevant, Eq.~(\ref{eq:ffr-num}) further supports indications from the
next-to-next-to-leading logarithmic threshold resummation~\cite{Vogt:2000ci,%
Catani:2003zt} that the corrections beyond the next-to-next-to-leading order
($\as^2$) of Refs.~\cite{Harlander:2002wh,Anastasiou:2002yz,Ravindran:2003um,%
Anastasiou:2005qj} are rather small.      

 
\subsection*{Acknowledgments}
 
The work of S.M. has been supported in part by the Helmholtz Gemeinschaft
under contract VH-NG-105 and by the Deutsche Forschungsgemeinschaft in
Sonderforschungs\-be\-reich/Transregio 9.
The work of J.V. has been part of the research program of the Dutch Foundation
for Fundamental Research of Matter (FOM).

\newpage


\end{document}